\documentclass[a4paper,fleqn,usenatbib]{mnras}

\pdfoutput=1

\usepackage{amssymb,amsmath,latexsym,mathrsfs,graphicx,subfigure,epsfig}
\usepackage{varioref,xr-hyper,multirow,array,diagbox,hyperref,wasysym}
\usepackage{color,float,mathtools,xcolor}
\usepackage[utf8]{inputenc}
\usepackage[T1]{fontenc}

\title[EDE and massive neutrinos]{Restoring cosmological concordance with early dark energy and massive neutrinos?}

\author[A. Reeves et al.]{
Alexander Reeves,$^{1}$\thanks{E-mail: \href{mailto:areeves@phys.ethz.ch}{areeves@phys.ethz.ch} (AR)}
Laura Herold,$^{2}$\thanks{E-mail: \href{mailto:lherold@mpa-garching.mpg.de}{lherold@mpa-garching.mpg.de} (LH)}
Sunny Vagnozzi,$^{3,4}$\thanks{E-mail: \href{mailto:sunny.vagnozzi@unitn.it}{sunny.vagnozzi@unitn.it} (SV)}
Blake D. Sherwin,$^{4,5}$\thanks{E-mail: \href{mailto:sherwin@damtp.cam.ac.uk}{sherwin@damtp.cam.ac.uk} (BDS)}
\newauthor
\hspace{0.7pt} and Elisa G. M. Ferreira$^{6,7}$\thanks{E-mail: \href{mailto:elisa.ferreira@ipmu.jp}{elisa.ferreira@ipmu.jp} (EGMF)} \\
$^{1}$Institute for Particle Physics and Astrophysics, ETH Z\"{u}rich, Wolfgang-Pauli-Stra\ss e 27, CH-8093 Z\"{u}rich, Switzerland \\
$^{2}$Max-Planck-Institut f\"{u}r Astrophysik, Karl-Schwarzschild-Stra\ss e 1, D-85740 Garching bei M\"{u}nchen, Germany \\
$^{3}$Department of Physics, University of Trento, Via Sommarive 14, 38123 Povo (TN), Italy \\
$^{4}$Kavli Institute for Cosmology, University of Cambridge, Madingley Road, Cambridge CB3 0HA, UK \\
$^{5}$Department of Applied Mathematics and Theoretical Physics, University of Cambridge, Wilberforce Road, Cambridge CB3 0WA, UK \\
$^{6}$Kavli IPMU (WPI), UTIAS, The University of Tokyo, 5-1-5 Kashiwanoha, Kashiwa, Chiba 277-8583, Japan \\
$^{7}$Instituto de F\'{i}sica, Universidade de S\~{a}o Paulo, Rua do Mat\~{a}o 1371, Butant\~{a}, 05508-090, S\~{a}o Paulo, Brazil
}

\date{Accepted XXX. Received YYY; in original form ZZZ}

\pubyear{2023}

\begin{document}
\label{firstpage}
\pagerange{\pageref{firstpage}--\pageref{lastpage}}
\maketitle

\begin{abstract}
\noindent The early dark energy (EDE) solution to the Hubble tension comes at the cost of an increased clustering amplitude that has been argued to worsen the fit to galaxy clustering data. We explore whether freeing the total neutrino mass $M_{\nu}$, which can suppress small-scale structure growth, improves EDE's fit to galaxy clustering. Using \textit{Planck} Cosmic Microwave Background and BOSS galaxy clustering data, a Bayesian analysis shows that freeing $M_{\nu}$ does not appreciably increase the inferred EDE fraction $f_{\rm EDE}$: we find the 95\%~C.L. upper limits $f_{\rm EDE}<0.092$ and $M_{\nu}<0.15\,{\rm eV}$.
Similarly, in a frequentist profile likelihood setting (where our results support previous findings that prior volume effects are important), we find that the baseline EDE model (with $M_{\nu}=0.06\,{\rm eV}$) provides the overall best fit. For instance, compared to baseline EDE, a model with $M_\nu=0.24\,{\rm eV}$ maintains the same $H_0$(km/s/Mpc)=(70.08, 70.11, respectively) whilst decreasing $S_8$=(0.837, 0.826) to the $\Lambda$CDM level, but worsening the fit significantly by $\Delta \chi^2=7.5$. For the datasets used, these results are driven not by the clustering amplitude, but by background modifications to the late-time expansion rate due to massive neutrinos, which worsen the fit to measurements of the BAO scale.
\end{abstract}


\section{Introduction}
\label{sec:intro}

The Hubble tension, i.e.\ the disagreement between independent measurements of the Hubble constant $H_0$, is arguably among cosmology's main open problems~\citep{DiValentino:2021izs,Perivolaropoulos:2021jda,Abdalla:2022yfr}. While systematics cannot be excluded~\citep{Freedman:2019jwv,Efstathiou:2020wxn,Mortsell:2021nzg}, serious consideration has been given to the possibility of new physics being at the origin of the tension, given its persistence~\citep{Mortsell:2018mfj,Guo:2018ans,Vagnozzi:2019ezj}. Consistency with Baryon Acoustic Oscillation (BAO) and uncalibrated SNeIa data requires new physics to preferably operate before recombination, in order to reduce the sound horizon by $\sim 7\%$~\citep{Bernal:2016gxb,Addison:2017fdm,Lemos:2018smw,Aylor:2018drw,Knox:2019rjx}.

One scenario invoked in this context is early dark energy (EDE), a model which introduces a pre-recombination dark energy (DE)-like component that boosts the expansion rate (reducing the sound horizon) before decaying~\citep{Poulin:2018cxd}. EDE fares well when confronted with Cosmic Microwave Background (CMB) and low-$z$ background data~\citep[see however][]{Krishnan:2020obg}, but was argued to be in tension with weak lensing (WL) and Large-Scale Structure (LSS) data~\citep{Hill:2020osr,Ivanov:2020ril,DAmico:2020ods}. It was hinted in \cite{Murgia:2020ryi,Smith:2020rxx} and shown in~\cite{Herold:2021ksg} that marginalization effects affect these analyses: a frequentist profile likelihood analysis found that large EDE fractions $f_{\rm EDE}$ are not ruled out by galaxy clustering data. However, parameter shifts in high $f_{\rm EDE}$ cosmologies lead to an increase in the clustering amplitude $\sigma_8$ and the related parameter $S_8$, worsening the ``$S_8$ discrepancy’’~\citep{DiValentino:2018gcu,Nunes:2021ipq}.

In this work, we study the influence of massive neutrinos on EDE, motivated by their free-streaming nature, whose associated power suppression might counteract the EDE-induced enhancement and provide a better fit to LSS data. We find no clear benefits for EDE resulting from massive neutrinos, neither in a Bayesian nor frequentist setting. We investigate prior volume effects, and physical effects driving our parameter constraints, which overall motivate further studies of EDE cosmologies with massive neutrinos.

\section{EDE and massive neutrinos}
\label{sec:edemnu}

The simplest EDE models envisage an ultra-light scalar field initially displaced from the minimum of its potential and frozen by Hubble friction, behaving as a DE component boosting the pre-recombination expansion rate.~\footnote{For examples of other EDE(-like) models, see~\cite{Karwal:2016vyq,Agrawal:2019lmo,Alexander:2019rsc,Lin:2019qug,Niedermann:2019olb,Ye:2020btb,Zumalacarregui:2020cjh,Gogoi:2020qif,Ballesteros:2020sik,Braglia:2020iik,Braglia:2020bym,Braglia:2020auw,Oikonomou:2020qah,Freese:2021rjq,Nojiri:2021dze,Karwal:2021vpk,Khosravi:2021csn,Niedermann:2021vgd,Sabla:2022xzj,Benevento:2022cql}.} Once the Hubble rate drops below its effective mass, the field becomes dynamical, rolls down and oscillates around the minimum of its potential. The canonical EDE model features a pseudoscalar (axion-like) field with the following potential:
\begin{eqnarray}
V(\phi) = m^2f^2 \left [ 1-\cos \left ( \frac{\phi}{f} \right ) \right ] ^n\,,
\label{eq:potential}
\end{eqnarray}
where $m$ and $f$ are the EDE mass and decay constant. With this choice of potential, EDE later decays as a fluid with effective equation of state $\langle w_{\phi} \rangle = (n-1)/(n+1)$.

The fundamental particle physics parameters $m$ and $f$ can be traded for the phenomenological parameters $f_{\rm EDE}$ and $z_c$: at redshift $z_c$, EDE's fractional contribution to the energy density is maximal and equal to $f_{\rm EDE} = \rho_{\rm EDE}/3M_{\rm Pl}^2H(z_c)^2$, where $\rho_{\rm EDE}$ is EDE's energy density, $M_{\rm Pl}$ is the \textit{Planck} mass, and $H(z)$ is the Hubble rate. The physics of the EDE model is then governed by four parameters: $f_{\rm EDE}$, $z_c$, $n$, and the initial misalignment angle $\theta_i = \phi_i/f$, with $\phi_i$ the initial field value. For simplicity we set $n=3$, corresponding to the best-fit value reported by~\cite{Poulin:2018cxd}. Increasing $f_{\rm EDE}$ reduces $r_{\rm drag}$, the sound horizon at the drag epoch, and solving the Hubble tension requires $f_{\rm EDE} \gtrsim 0.1$.

\begin{figure}
\centering
\includegraphics[width=1.0\linewidth]{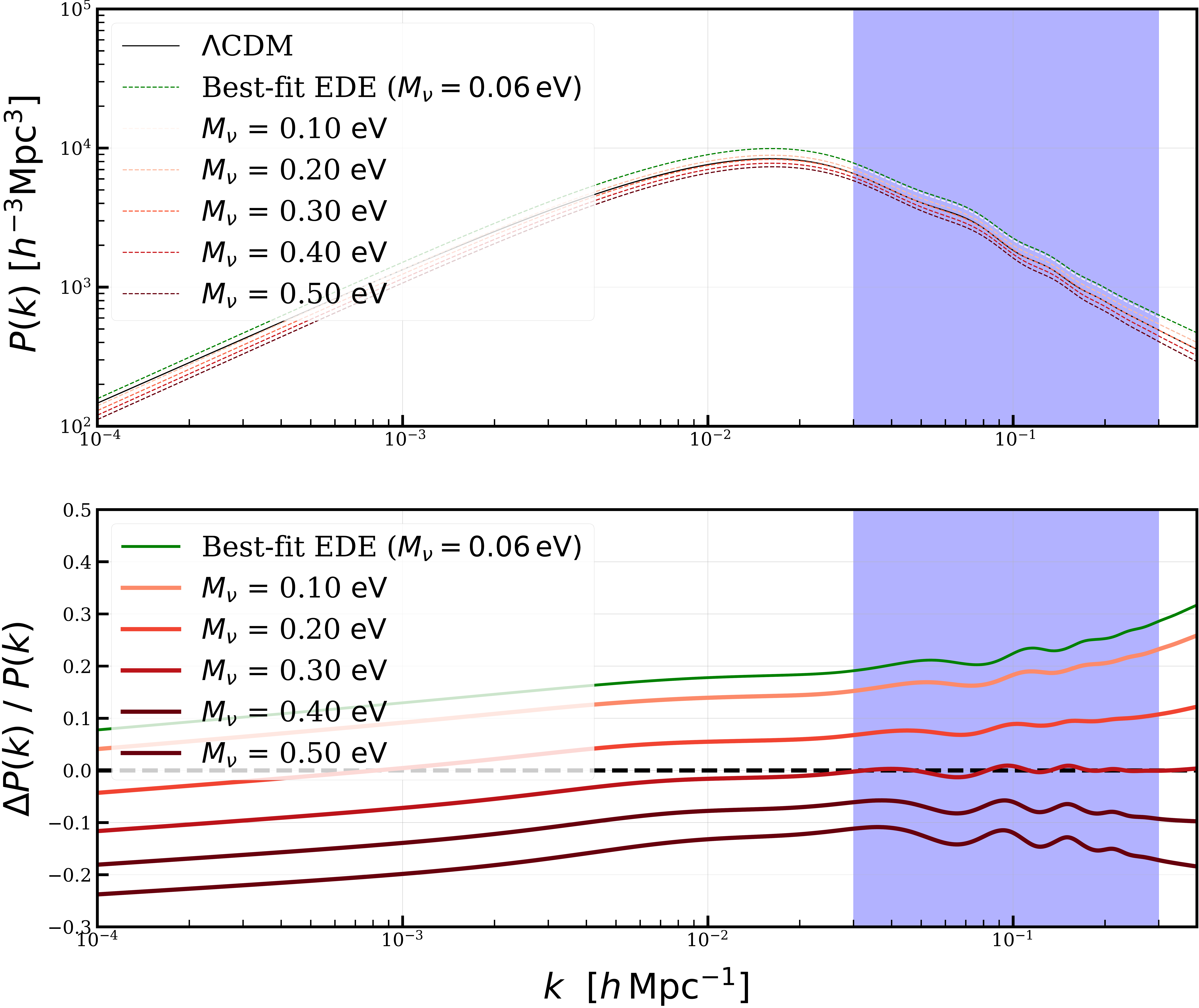}
\caption{Impact of $M_{\nu}$ on the EDE matter power spectrum, with the other parameters (including $\theta_s$ and nuisance parameters) fixed to the best-fit values of~\citet{Hill:2020osr}. \textit{Lower panel}: relative change with respect to $\Lambda$CDM. The purple region is the wavenumber range of interest to current surveys.}
\label{fig:mnu_ede_pk}
\end{figure}

To compensate for the EDE-induced enhancement of the early integrated Sachs-Wolfe (eISW) effect and preserve the fit to the CMB~\citep{Vagnozzi:2021gjh}, EDE's success comes at the significant cost of an increase in the dark matter (DM) density $\omega_c=\Omega_c h^2$. This boosts the matter power spectrum and raises $S_8 \propto \sigma_8\sqrt{\Omega_m}$, worsening the $S_8$ discrepancy present within $\Lambda$CDM (see Fig.~\ref{fig:mnu_ede_pk}). EDE was thus argued to be disfavored by WL and galaxy clustering data~\citep{Hill:2020osr}, although ~\cite{Murgia:2020ryi},~\cite{Smith:2020rxx},~\cite{Herold:2021ksg}, and~\cite{Gomez-Valent:2022hkb} argued that this is in part due to prior volume effects (PVEs).~\footnote{In the above, the CMB data is from \textit{Planck}. Mild preferences for EDE have been found from ACT or SPT data, or dropping \textit{Planck} high-$\ell$ data~\citep{Hill:2021yec,Chudaykin:2020acu,Jiang:2021bab,Poulin:2021bjr,LaPosta:2021pgm,Jiang:2022uyg,Ye:2022efx,Jiang:2022qlj}, but consensus on these results is lacking, due to possible systematics~\citep[e.g.][]{Handley:2020hdp,Smith:2022hwi}.}

A possible remedy is to add extra components absorbing the excess power~\citep[e.g.][]{Allali:2021azp,Ye:2021iwa,Clark:2021hlo}. Massive neutrinos are an economical and conservative candidate in this sense as we know oscillation experiments show that at least two neutrino mass eigenstates are massive. Including a free neutrino mass sum $M_{\nu}$ (rather than fixing it to the minimum allowed value of $0.06\,{\rm eV}$ as in baseline EDE) can thus be justified invoking only known physics and this inclusion has not been explored in this context so far. Due to their free-streaming nature, massive neutrinos suppress small-scale power~\citep{Lesgourgues:2006nd}: Fig.~\ref{fig:mnu_ede_pk} shows how values of $M_{\nu} \approx 0.3\,{\rm eV}$ can in principle absorb the EDE-induced excess power in a wavenumber range relevant to current surveys. Note that models connecting EDE to neutrinos and predicting high $M_{\nu}$ have been studied~\citep{Sakstein:2019fmf,CarrilloGonzalez:2020oac}, alongside the role of neutrino physics in relation to cosmic tensions~\citep{Ilic:2019pwq,Das:2021pof,DiValentino:2021imh,Sakr:2021jya,Chudaykin:2022rnl}.

Adding $M_{\nu}$ as a free parameter within $\Lambda$CDM induces well-known parameter degeneracies at the CMB level: a negative $M_\nu$-$H_0$ correlation related to the geometrical degeneracy, and a positive $M_\nu$-$\omega_c$ correlation connected to the CMB lensing amplitude~\citep{Vagnozzi:2018jhn,RoyChoudhury:2019hls}. BAO data partially aid in breaking these degeneracies (especially the $M_{\nu}$-$H_0$ one). At fixed acoustic scale $\theta_s$, increasing $M_\nu$ reduces the BAO angular scale $\theta_{\rm BAO} = r_{\rm drag}/D_V(z_{\rm eff})$~\citep{Hou:2012xq,Archidiacono:2016lnv,Boyle:2017lzt}, with $D_V(z_{\rm eff})$ the volume-averaged distance at the effective redshift $z_{\rm eff}$.

\section{Datasets and methodology} 
\label{sec:datasets}

We use \textit{Planck} 2018 CMB temperature, polarization, and lensing measurements, combining the \texttt{Plik TTTEEE}, \texttt{lowl}, \texttt{lowE}, and \texttt{lensing} likelihoods~\citep{Planck:2019nip}. We add the joint pre-reconstruction full-shape (FS) plus post-reconstruction BAO likelihood for the BOSS DR12 galaxies~\citep[see][]{Ivanov:2019pdj,Philcox:2020vvt}.\footnote{In future work we will study the impact of updates in the modeling of the window function~\citep{Beutler:2021eqq}. We do not expect a big impact on our constraints, which are driven by the BAO scale.} The cross-covariance between FS and BAO is fully taken into account in the likelihood. The FS measurements include both the monopole and quadrupole moments. We do not include a distance ladder $H_0$ prior to not bias $H_0$ towards high values~\citep[see also][]{Efstathiou:2021ocp}. 

We consider a 10-parameter EDE+$M_{\nu}$ model where, besides the 6 $\Lambda$CDM parameters, $M_{\nu}$ and 3 EDE parameters ($f_{\rm EDE}$, $\log_{10}z_c$, and $\theta_i$, fixing $n=3$) are varied. The neutrino mass spectrum is modelled following the degenerate approximation, sufficiently accurate for the precision of current data~\citep{Vagnozzi:2017ovm,Giusarma:2018jei,RoyChoudhury:2019hls,Archidiacono:2020dvx,Tanseri:2022zfe}. For comparison, we also consider 3 related models: 9-parameter EDE ($M_{\nu}=0.06\,{\rm eV}$), 7-parameter $\Lambda$CDM+$M_{\nu}$ ($f_{\rm EDE}=0$), and the standard 6-parameter $\Lambda$CDM.

Theoretical predictions are computed using the \texttt{EDE-CLASS-PT} Boltzmann solver\footnote{\url{https://github.com/Michalychforever/EDE_class_pt}}, itself a merger of \texttt{CLASS\_EDE}~\citep{Hill:2020osr} and \texttt{CLASS-PT}~\citep{Philcox:2020vvt}, themselves both extensions to the Boltzmann solver \texttt{CLASS}~\citep{Blas:2011rf}. The underlying galaxy power spectrum model is based on the Effective Field Theory of LSS~\citep[EFTofLSS,][]{Baumann:2010tm}, which is the most general, symmetry-driven model for the mildly non-linear clustering of biased tracers of the LSS, accounting for the complex and poorly-known
details of short-scale physics which are integrated out.

We follow two analysis methods. We begin with a standard Bayesian analysis, adopting Monte Carlo Markov Chain (MCMC) methods and using the \texttt{MontePython} MCMC sampler~\citep{Audren:2012wb,Brinckmann:2018cvx}. We impose the same (flat) priors on the EDE parameters as in~\cite{Hill:2020osr}, whereas for the EFTofLSS nuisance parameters we follow~\cite{Philcox:2020vvt}. We monitor the convergence of the generated MCMC chains via the Gelman-Rubin parameter $R-1$~\cite{Gelman-Rubin}, with the chains considered to be converged if $R-1<0.05$ (which, we note, is a more stringent requirement than that adopted by several other EDE works). Following the conclusions of~\cite{Herold:2021ksg,Herold:2022iib}, and the analysis in~\cite{Planck:2013nga} for varying neutrino mass sum, we then perform a profile likelihood (PL) analysis in $M_{\nu}$: for a given (fixed) value of $M_{\nu}$, after minimizing the $\chi^2$ with respect to all other parameters, the PL is given by $\Delta \chi^2(M_{\nu})$. We follow the minimization method of~\cite{Schoneberg:2021qvd}, referred to as S21, running a series of MCMCs with decreased temperature and enhanced sensitivity to likelihood differences. For comparison we also use the gradient descent-based \texttt{Migrad} algorithm~\citep{James:1975dr}, finding that S21 always outperforms it for the EDE model.

\section{Results}
\label{sec:results}

\begin{figure*}
\centering
\includegraphics[width=1.0\linewidth]{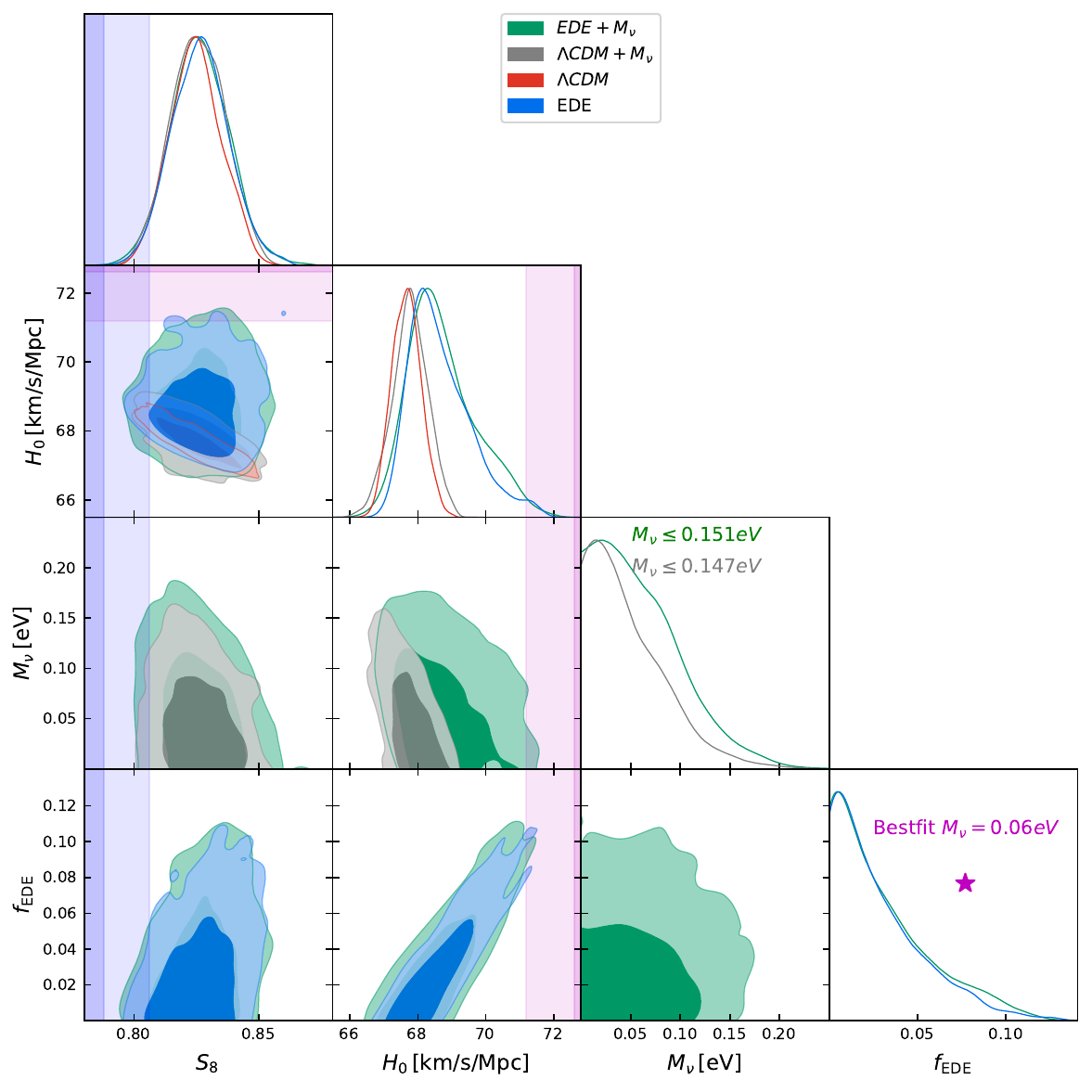}
\caption{1D and 2D posteriors for $S_8$, $H_0$, $f_{\rm EDE}$ and $M_{\nu}$ within different models (see color coding). These contours represent the Bayesian constraints obtained when combining \textit{Planck} and BOSS (FS+BAO) data. Pink bands indicate the SH0ES local $H_0$ measurement from~\citet{Riess:2021jrx}, and purple bands denote the inverse-variance-weighed combination of \textit{DES-Y1+KiDS+HSC} $S_8$ measurements as in~\citet{Hill:2020osr}. The best-fit $f_{\rm EDE}$ value with fixed $M_\nu=0.06\,{\rm eV}$ is shown as a purple star.}
\label{fig:triangular_ede_mnu}
\end{figure*}

\begin{figure}
\centering
\includegraphics[width=1.0\linewidth]{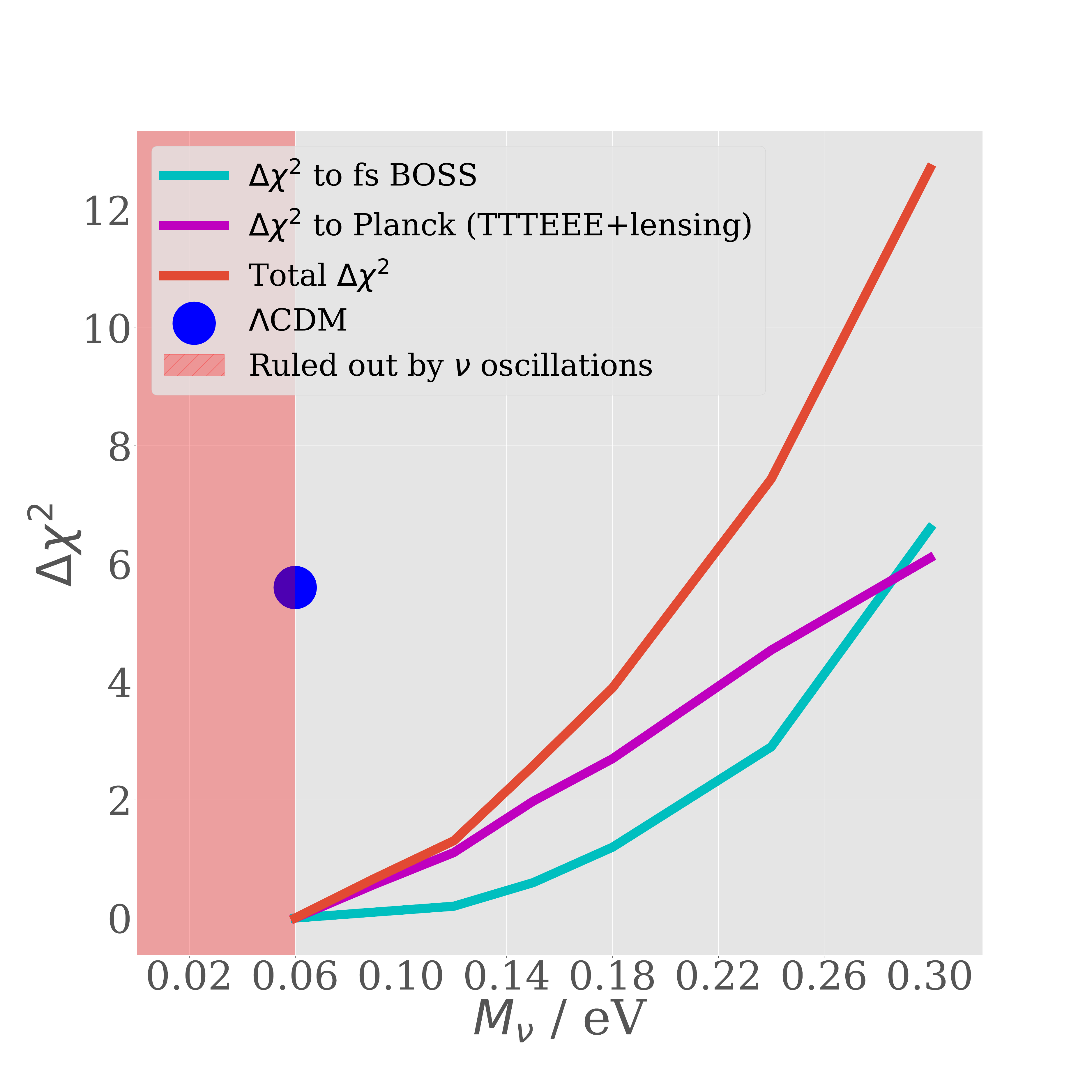}
\caption{$\chi^2$ contributions as a function of $M_\nu$ within the EDE model. The purple and blue lines respectively show the $\chi^2$ contribution from the \textit{Planck} and BOSS likelihoods and the red line is the total $\chi^2$, given by the sum of the two. The blue dot represents the best-fit $\Lambda$CDM model, given the same combination of data. The red shaded region encompasses values of $M_\nu$ which are ruled out by oscillation experiments. The full table of best-fit results is shown in Appendix~A.}
\label{fig:chi2_breakdown}
\end{figure}

\begin{figure}
\centering
\includegraphics[width=1.0\linewidth]{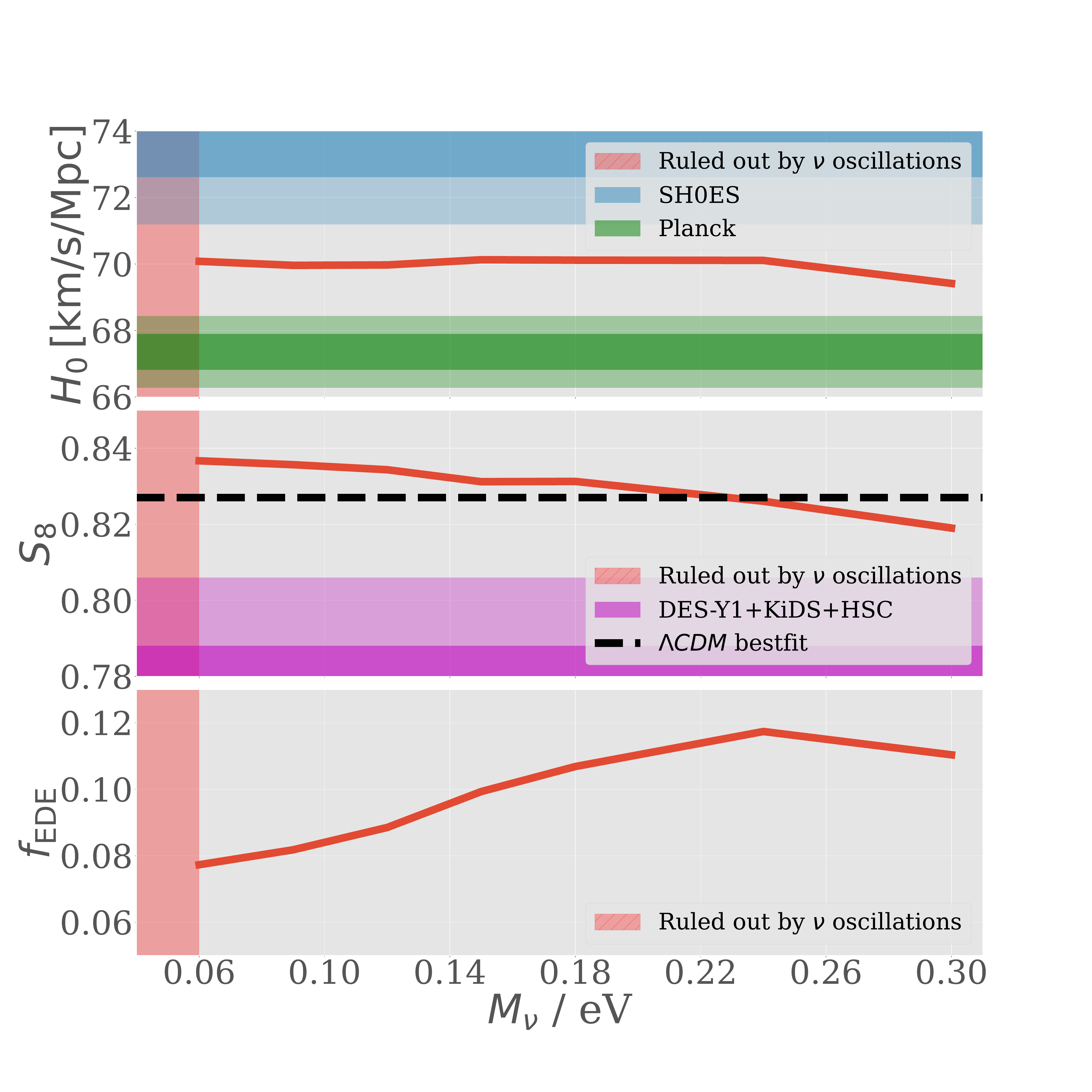}
\caption{Variation in the best-fit values of selected cosmological parameters as a function of $M_\nu$. The red shaded region encompasses values of $M_\nu$ that are ruled out by oscillation experiments. The blue and green bands indicate respectively the value of $H_0$ inferred from \textit{Planck} assuming the $\Lambda$CDM model~\citep{Planck:2018vyg}, and the SH0ES local distance ladder value~\citep{Riess:2021jrx}. The purple band is an inverse-variance-weighed combination of DES-Y1+KiDS+HSC $S_8$ measurements as in~\citet{Hill:2020osr}, whilst the black dashed line is the best-fit value of $S_8$ from a fit to the same datasets assuming $\Lambda$CDM. The full table of best-fit values is shown in Appendix~\ref{sec:appendix}.}
\label{fig:param_shifts}
\end{figure}

From the \textit{Planck}+BOSS combination, a Bayesian analysis of the EDE+$M_{\nu}$ model returns the 95\% confidence level (C.L.) upper limit $M_{\nu}<0.151\,{\rm eV}$. This is only slightly weaker than the corresponding $\Lambda$CDM+$M_{\nu}$ limit from the same dataset ($M_{\nu}<0.147\,{\rm eV}$), safely excluding the ballpark region required to compensate the EDE enhancement ($M_{\nu} \sim 0.3\,{\rm eV}$). This reflects in sub-$\sigma$ shifts and slightly broader uncertainties in $H_0$, $\sigma_8$, and $f_{\rm EDE}$, compared to their baseline EDE ($M_{\nu}=0.06\,{\rm eV}$) counterparts [in brackets]: $H_0=68.71 \pm 1.06 $ $[68.72 \pm 0.90]\,{\rm km}/{\rm s}/{\rm Mpc}$, $S_8=0.826 \pm 0.012$ $[0.826 \pm 0.012]$, $f_{\rm EDE}<0.092$ $[<0.085]$, see also Fig.~\ref{fig:triangular_ede_mnu}. These sub-$\sigma$ shifts show that, in a Bayesian setting, freeing $M_{\nu}$ does not significantly increase the inferred $f_{\rm EDE}$, with the peak of the posterior still being close to zero.

We then perform a PL analysis, fixing $M_{\nu}$ to seven values between $0.06\,{\rm eV}$ and $0.3\,{\rm eV}$ and dissecting each likelihood's contribution to the total $\chi^2$. We aim to identify \textit{a)} which dataset(s) prevent high $M_{\nu}$ values, and \textit{b)} whether PVEs are playing a role. \cite{Smith:2019ihp},~\cite{Herold:2021ksg}, and~\cite{Gomez-Valent:2022hkb} argued that PVEs play a key role with EDE, as in the $f_{\rm EDE} \to 0$ limit $\Lambda$CDM is recovered, so the likelihood is approximately flat in the $\theta_i$ and $z_c$ directions. This leads to a larger prior volume in the low $f_{\rm EDE}$ region, resulting in a preference for small $f_{\rm EDE}$ upon marginalization. The PL is not impacted by these PVEs.

Our PL analysis results are shown in Fig.~\ref{fig:chi2_breakdown} and Fig.~\ref{fig:param_shifts}. We find that the baseline EDE model ($M_{\nu} = 0.06\,{\rm eV}$) with $f_{\rm EDE} = 0.077$ fits the data best. This has a $\Delta \chi^2=-5.6$ compared to the baseline $\Lambda$CDM model although we have introduced three extra parameters (when fixing $M_\nu$). Following~\citet{1100705}, we can compute the Akaike information criterion (AIC), a measure of statistical preference for models. It accounts for a differing number of free parameters, penalizing a higher number of free parameters, which does not lead to a sufficient improvement in fit. For a given model it is given by:
\begin{eqnarray}
\text{AIC} = 2k + \min(\chi^2)
\end{eqnarray}
where $k$ is the number of model parameters, and where a lower AIC indicates a model which is statistically preferred. For the EDE model with $M_\nu=0.06\,{\rm eV}$ we find $\Delta {\rm AIC}=+0.4$ compared to $\Lambda$CDM, indicating a mild statistical preference for $\Lambda$CDM despite the overall reduction in $\chi^2$. The best-fit $f_{\rm EDE}$ for this model is significantly higher than the mean value expected from the Bayesian results for the baseline model with $M_\nu=0.06\,{\rm eV}$ (see also the purple star in Fig.~\ref{fig:triangular_ede_mnu}) hence we reconfirm the results of~\citet{Herold:2021ksg,Gomez-Valent:2022hkb} that PVEs could have an impact on the Bayesian constraints of the baseline EDE model. However, even once this effect is accounted for in the PL analysis, there is no evidence of benefit from a raised $M_\nu$ in the EDE scenario. Lowering $S_8$ to the $\Lambda$CDM level within EDE requires $M_{\nu} \sim 0.24\,{\rm eV}$ ($S_8=0.826$, $f_{\rm EDE}=0.117$). This comes at the cost of a substantially worse fit quality ($\Delta \chi^2=7.5$), clearly disfavouring this model.

The profile likelihood in $M_\nu$, broken down into the $\chi^2$ contributions from the individual datasets in our analysis is shown in the blue and purple lines in Fig.~\ref{fig:chi2_breakdown} (related information is shown in Fig.~\ref{fig:param_shifts}). We find that the fit to both the \textit{Planck} TTTEEE + lensing and the BOSS data worsens as $M_\nu$ is increased. For the \textit{Planck} data the strong constraining power on $M_\nu$ is  expected (\cite{Planck:2018vyg} for $\Lambda$CDM). More interestingly, the fit to the BOSS dataset also degrades monotonically with $M_\nu$: this suggests that the benefits of increased $M_\nu$ in the EDE scenario in terms of a reduction in clustering amplitude are being outweighed by an increasing mismatch to the geometric features of the FS spectrum. We find that most of the effect of EDE-induced parameter shifts and $M_{\nu}$ on the FS clustering amplitude is re-absorbed by nuisance parameter shifts, as pointed out in~\cite{Ivanov:2020ril} within baseline EDE. The remaining differences in the galaxy power spectrum multipoles are due to a mismatch in the location of the BAO wiggles. Hence, the derived constraints on the EDE+$M_{\nu}$ model are mostly driven by shifts in the BAO scale $\theta_{\rm BAO}$, rather than the $M_{\nu}$-driven small-scale power suppression (see further discussion in Appendix~\ref{sec:appendix2}). In Fig.~\ref{fig:baopos} we show how the fit to the BAO scale gradually worsens as $M_{\nu}$ increases, reflecting the increasing trend in the BOSS likelihood $\chi^2$.

The increase in $M_{\nu}$ is accompanied by different parameter shifts as demonstrated in Fig.~\ref{fig:param_shifts}. We find a $M_{\nu}$-$f_{\rm EDE}$ correlation which can be understood as follows. Increasing $M_{\nu}$ at fixed $\theta_s$ and $\omega_b+\omega_c$ results in the $z \lesssim 1$ expansion rate decreasing relative to a $M_{\nu}=0$ model~\citep[see a complete explanation in][]{Hou:2012xq,Archidiacono:2016lnv}, decreasing $\theta_{\rm BAO}$. In contrast, raising $f_{\rm EDE}$ leads to a fractional decrease in $r_{\rm drag}$ which, as a result of the accompanying increase in $H_0$, results in a larger fractional decrease in $D_V(z_{\rm eff})$. The overall effect is to (re-)increase $\theta_{\rm BAO}$, as we checked numerically. The net result is that $\theta_{\rm BAO}$ still decreases when increasing $M_{\nu}$ and $f_{\rm EDE}$ simultaneously, but less so than if we had kept $f_{\rm EDE}$ fixed. The extent to which $f_{\rm EDE}$ can compensate for the $M_{\nu}$-induced reduction of $\theta_{\rm BAO}$ is strongly limited by the accompanying increase in $\omega_c$ (compensating the eISW boost), whose effect is similar to that of raising $M_{\nu}$, overall (re-)decreasing $\theta_{\rm BAO}$. As a result, the best-fit $H_0$ barely shifts when $M_{\nu}$ is raised. These arguments easily extend to anisotropic BAO measurements~\citep[see also][]{Klypin:2020tud}.See~\cite{Lattanzi:2017ubx,Vagnozzi:2019utt,Sakr:2022ans} for more complete discussions on the effect of massive neutrinos on various cosmological probes.

For $M_{\nu} \gtrsim 0.18\,{\rm eV}$ the $\chi^2$ increases more steeply, mostly driven by the BOSS likelihood due to the gradually worsened BAO scale fit. However, $H_0$ remains stable within $1\%$ across the whole $M_{\nu}$ range, due to two competing effects: while increasing $f_{\rm EDE}$ pulls $H_0$ upwards, increasing $M_\nu$ lowers it due to the geometrical degeneracy. As discussed earlier, increasing $M_{\nu}$ is accompanied by decreases in $\sigma_8$ and $S_8$.


\begin{figure}
\centering
\includegraphics[width=1.0\linewidth]{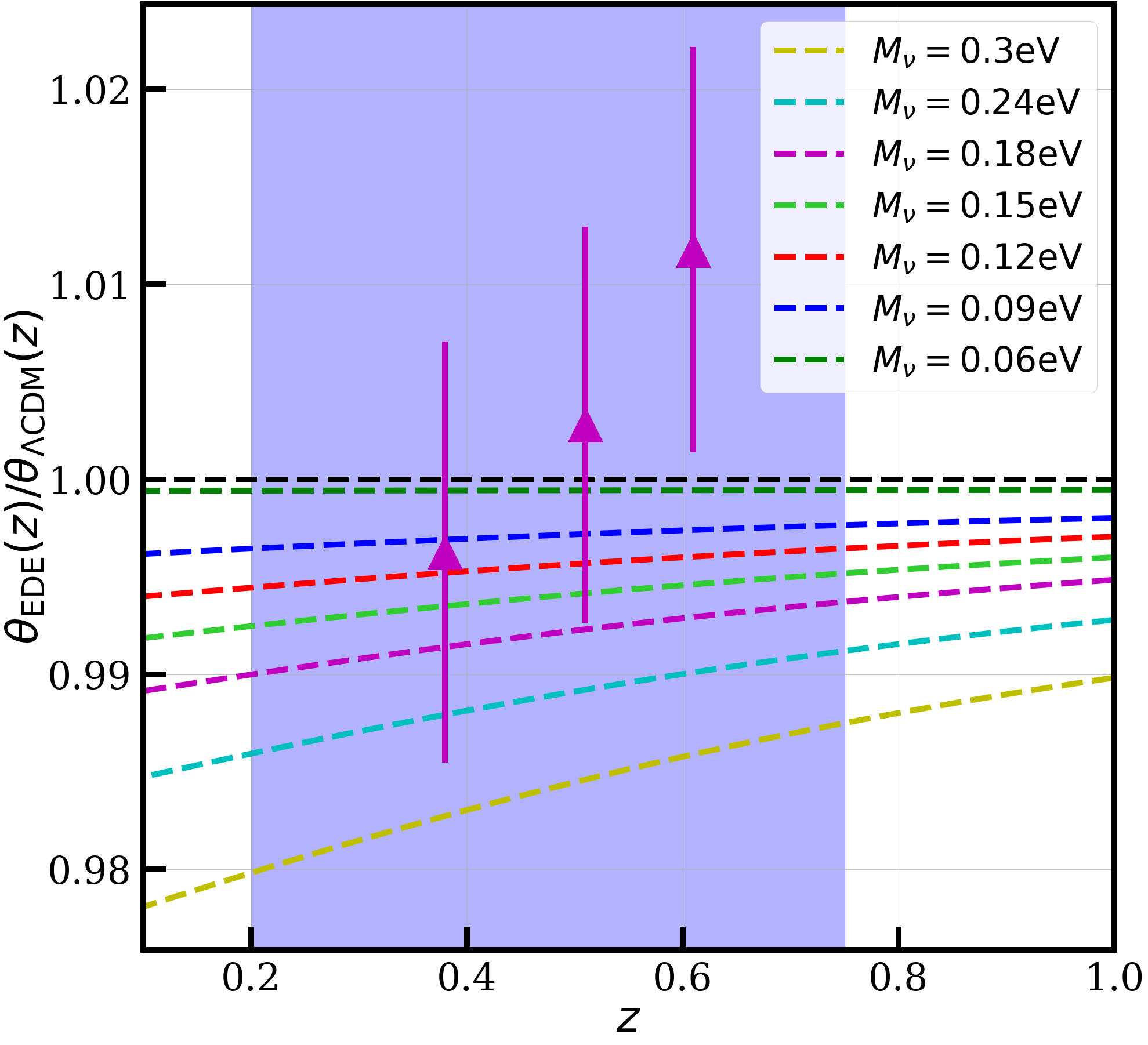}
\caption{BAO angular scale within EDE at fixed values of $M_{\nu}$ relative to the $\Lambda$CDM predictions (all parameters fixed to their \textit{Planck}+BOSS best fits). Purple triangles denote the BOSS DR12 consensus isotropic BAO measurements~\citep{BOSS:2016wmc}.}
\label{fig:baopos}
\end{figure}

\section{Conclusions} 
\label{sec:conclusions}

It is well known that introducing early dark energy (EDE) in order to resolve the $H_0$ tension worsens the ``$S_8$ tension''. Our paper re-examines this issue in light of an extension including massive neutrinos, driven by the possibility of their small-scale power suppression counteracting the EDE-induced excess power, which leads to the increase in $S_8$. 

A standard Bayesian analysis of CMB and galaxy clustering data shows that freeing $M_{\nu}$ does not increase the inferred $f_{\rm EDE}$, and has no effect on EDE’s standings relative to the $H_0$ and $S_8$ tensions. A frequentist profile likelihood analysis also finds no clear benefits for EDE resulting from a higher $M_{\nu}$, as the best fit is achieved within baseline EDE ($M_{\nu}=0.06\,{\rm eV}$), but supports earlier claims of PVEs playing a key role in these Bayesian constraints using BOSS data~\citep{Smith:2020rxx,Herold:2021ksg,Gomez-Valent:2022hkb}. Values of $M_{\nu}$ lowering $S_8$ to the $\Lambda$CDM level are not preferred statistically; a model with $M_\nu=0.24\,{\rm eV}$ worsens the fit by $\Delta \chi^2 = 7.5$ in comparison to baseline EDE. We find a correlation between $f_{\rm EDE}$ and $M_{\nu}$, along with the expected negative $M_{\nu}$-$S_8$ correlation.~\footnote{As a caveat, we note that the perturbation theory and mode-coupling kernels used in \texttt{CLASS-PT} have been computed assuming an Einstein-de Sitter Universe, whereas here we are including both EDE and neutrino masses: as these new physics contributions do not violate the equivalence principle, this is a reasonable approximation (although one that would need to be refined for future more precise data), see e.g. more complete recent discussions in Sec.~IVF of~\cite{Chudaykin:2020aoj} and Sec.~IIB of~\cite{Nunes:2022bhn}, with similar considerations holding for the IR resummation procedure.}

Contrary to initial expectations, our $M_{\nu}$ limits are driven not by the full-shape clustering amplitude (re-absorbed by nuisance parameters), but by shifts in the BAO scale $\theta_{\rm BAO}$. As the clustering amplitude plays a minor role, our analysis is not very sensitive to the benefits of the $M_{\nu}$-driven power suppression. One possible avenue for further work would be to explore the inclusion of WL data or WL-derived priors which, without freeing $M_{\nu}$, appear to slightly decrease the value of $f_{\rm EDE}$ and consequently $H_0$~\citep{Herold:2022iib}; it will be interesting to study whether freeing $M_{\nu}$ can improve the consistency of EDE with WL measurements. A related recent paper by some of us, which appeared after ours was posted on arXiv, has derived new PL-based confidence intervals on EDE using additional datasets~\citep[including a Gaussian likelihood centered on the $S_8$ of the Dark Energy Survey Year
3 analysis, see][]{Herold:2022iib}.

In the coming years, $\beta$-decay experiments will aim for a model-independent kinematical neutrino mass detection which, combined with future cosmological probes~\citep{SimonsObservatory:2018koc,SimonsObservatory:2019qwx}, will set the stage for further tests of EDE and massive neutrinos.

\section*{Acknowledgements}

We thank George Efstathiou, Colin Hill, Eiichiro Komatsu and Oliver Philcox for many useful discussions. S.V. was partially supported by the Isaac Newton Trust and the Kavli Foundation through a Newton-Kavli Fellowship, and by a grant from the Foundation Blanceflor Boncompagni Ludovisi, n\'{e}e Bildt. B.D.S. is supported by the European Research Council (Grant agreement No.~851274) and an STFC Ernest Rutherford Fellowship. The Kavli IPMU is supported by World Premier International Research Center Initiative (WPI), MEXT, Japan.

\section*{Data availability}

The data underlying this article will be shared upon request to the corresponding author(s).

\section*{Appendix A: Frequentist table}
We present the full table of frequentist results considering the combination of \textit{Planck} and BOSS data. Some of this information is displayed graphically in Fig.~\ref{fig:chi2_breakdown} and Fig.~\ref{fig:param_shifts}. 
\label{sec:appendix}
\begin{table*}
\begin{center}
{\begin{tabular}{|c||cccccccc|}
\hline \hline
\multicolumn{9}{|c|}{\textbf{Individual best-fit} $\boldsymbol{\chi^2}$ \textbf{contributions}} \\ \hline
\backslashbox{Likelihood}{Model} & $\Lambda$CDM$_{0.06}$ & EDE$_{0.06}$ & EDE$_{0.09}$ & EDE$_{0.12}$ & EDE$_{0.15}$ & EDE$_{0.18}$ & EDE$_{0.24}$ & EDE$_{0.3}$ \\ \hline
BOSS (BAO+FS) & 297.2 & 295.3 & 295.4 & 295.5 & 295.9 & 296.5 & 298.2 & 301.9 \\
\textit{Planck} \texttt{TTTEEE} & 2345.5 & 2342.6 & 2343.2 & 2343.7 & 2345.1 &2345.5 & 2347.2 &  2348.3 \\
\textit{Planck} \texttt{lowE} & 396.3 & 396.1 & 396.4 & 396.8 & 396.5& 397.0 & 397.3 & 397.7 \\
\textit{Planck} \texttt{lowl} & 23.2 & 21.9 & 21.7 & 21.5 & 21.3 & 21.2 &  21.1 & 21.1 \\
\textit{Planck} \texttt{lensing} & 8.8  & 9.47 & 9.34 &9.18 &  9.15 & 9.07 & 9.01 & 9.07 \\ \hline
Total $\chi^2$ (S21) & 3071.0 &  3065.4 & 3065.9 & 3066.7 & 3067.9 & 3069.3 & 3072.9 & 3078.1 \\ 
(\texttt{Migrad}) & 3078.6 & 3070.7 & 3072.7 &3073.0 &3073.4 &3076.0 &3076.5 &3088.3  \\ \hline \hline
\multicolumn{9}{|c|}{\textbf{Best-fit parameters}} \\ \hline
$H_0\,[{\rm km}/{\rm s}/{\rm Mpc}]$ & 67.59 & 70.08 & 69.96 & 69.97 & 70.12 & 70.12  & 70.11 & 69.42 \\
$\sigma_8$ & 0.811 & 0.828 & 0.824 & 0.820 & 0.814 & 0.811 &  0.802 & 0.787	\\
$\Omega_m$ & 0.312 & 0.306 & 0.309 & 0.311 & 0.312 & 0.315 & 0.319 & 0.325 \\
$S_8$ &  0.827 & 0.837 & 0.836 & 0.834 & 0.831 & 0.831 & 0.826 & 0.819 \\
$\omega_c$ & 0.120 & 0.127 & 0.128 & 0.128 & 0.129 & 0.130 & 0.131 & 0.130 \\
$f_{\rm EDE}$ & -- & 0.077 & 0.082 & 0.089 & 0.099 & 0.107 & 0.117 & 0.117 \\
\hline \hline
\end{tabular}}
\end{center}
\caption{\textit{Upper half}: breakdown of the best-fit $\chi^2$ contributions from each likelihood and the total best-fit $\chi^2$, within different models (``EDE$_x$'' indicates an EDE model with fixed $M_{\nu}=x\,{\rm eV}$). \textit{Lower half}: best-fit values of $H_0$, $\sigma_8$, $\Omega_m$, $S_8$, $\omega_c$ and $f_{\rm EDE}$ within each model.}  
\label{tab:ind_chi}
\end{table*}
The full set of frequentist results showing the breakdown of the $\chi^2$ and parameter shifts is shown in Tab.~\ref{tab:ind_chi}. The baseline results for this work were produced following the minimisation routine of \citet{Schoneberg:2021qvd}. We checked that \texttt{Migrad} recovers a similar trend, albeit with $\chi^2$ values consistently higher than S21. 

\section*{Appendix B: Data comparisons}
\label{sec:appendix2}

\begin{figure*}
\centering
\includegraphics[width=0.8\linewidth]{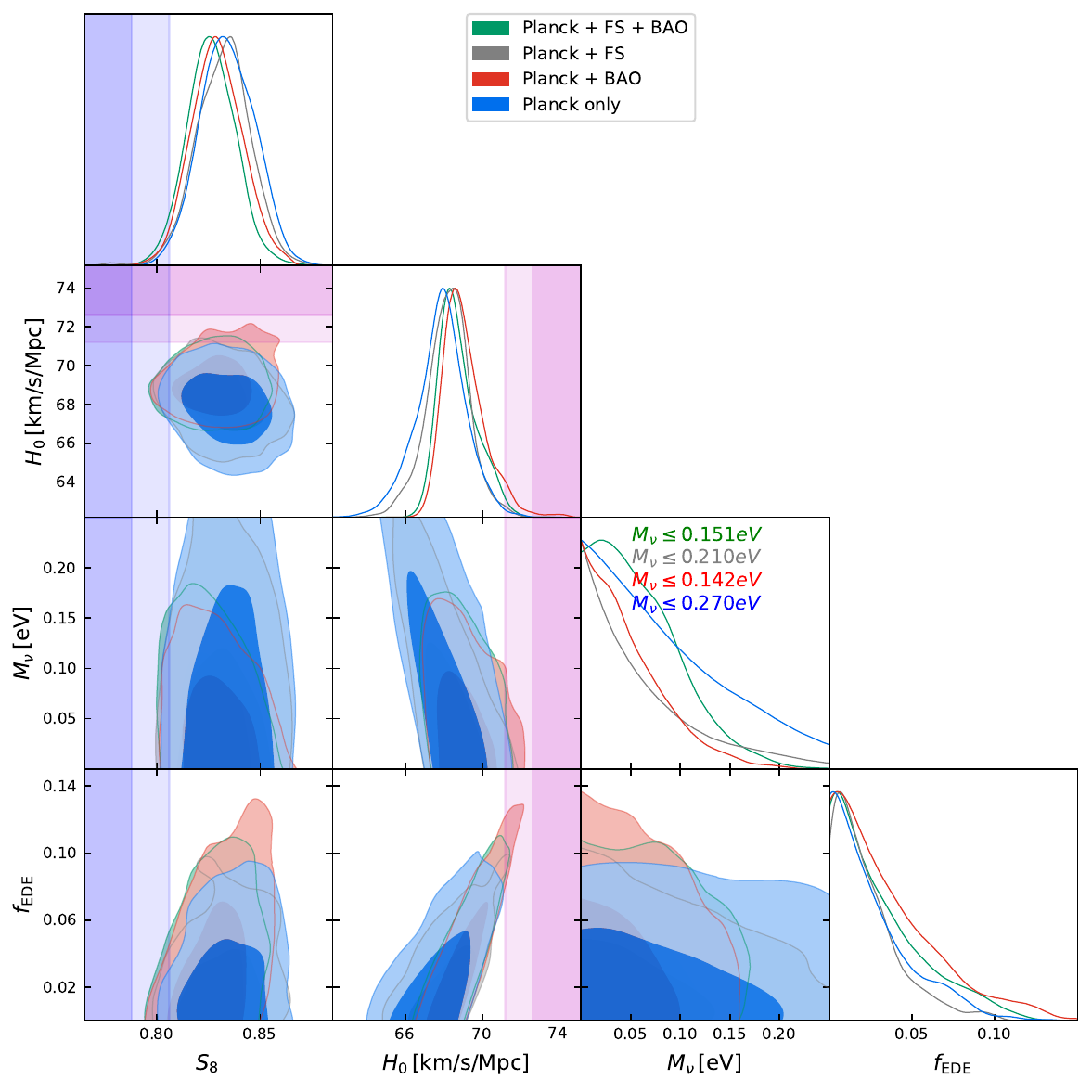}
\caption{MCMC contours for the EDE+$M_{\nu}$ model obtained from several combinations of BOSS (FS and/or BAO) and \textit{Planck} data.}
\label{fig:datacheck}
\end{figure*}

We checked how different combinations of BOSS data affect the results presented in this analysis. Fig.~\ref{fig:datacheck} shows corner plots for different combinations of the datasets we used. There is a clear gain in the constraining power of the data on $M_{\nu}$ when moving from \textit{Planck} alone (blue) to any of the contours that contain BOSS data in addition. However, there is little difference between the \textit{Planck}+\textit{BAO} and \textit{Planck}+\textit{BAO}+\textit{FS} constraints, confirming earlier results in the literature~\citep{Ivanov:2019hqk}. The most stringent constraint on $M_\nu$ is obtained when in addition to \textit{Planck} data we consider the post-reconstruction BAO likelihood ($M_\nu < 0.144\,{\rm eV}$), which suggests that geometric features in BOSS data are what drives the constraints in the full FS+BAO likelihood for which we find $M_\nu < 0.151\,{\rm eV}$ (on the other hand from the \textit{Planck}+FS combination we find the looser constraint $M_\nu < 0.210\,{\rm eV}$). These results all agree with earlier findings in the literature~\citep[see e.g.][]{Ivanov:2019hqk,Tanseri:2022zfe}, confirming that the constraining power for $M_\nu$ of BOSS data is mostly contained in the geometrical, rather than shape information. This explains the marginal role the amplitude of clustering (as opposed to the position of the BAO peaks) appears to play in our $M_{\nu}$ constraints, as discussed throughout the paper. Finally, it is worth pointing out that the FS and combined FS+BAO likelihoods feature seven additional EFTofLSS nuisance parameters compared to the BAO-only likelihood.

\bibliographystyle{mnras}
\bibliography{EDE_massive_neutrinos.bib}

\begin{thebibliography}{}
\makeatletter
\relax
\def\mn@urlcharsother{\let\do\@makeother \do\$\do\&\do\#\do\^\do\_\do\%\do\~}
\def\mn@doi{\begingroup\mn@urlcharsother \@ifnextchar [ {\mn@doi@}
  {\mn@doi@[]}}
\def\mn@doi@[#1]#2{\def\@tempa{#1}\ifx\@tempa\@empty \href
  {http://dx.doi.org/#2} {doi:#2}\else \href {http://dx.doi.org/#2} {#1}\fi
  \endgroup}
\def\mn@eprint#1#2{\mn@eprint@#1:#2::\@nil}
\def\mn@eprint@arXiv#1{\href {http://arxiv.org/abs/#1} {{\tt arXiv:#1}}}
\def\mn@eprint@dblp#1{\href {http://dblp.uni-trier.de/rec/bibtex/#1.xml}
  {dblp:#1}}
\def\mn@eprint@#1:#2:#3:#4\@nil{\def\@tempa {#1}\def\@tempb {#2}\def\@tempc
  {#3}\ifx \@tempc \@empty \let \@tempc \@tempb \let \@tempb \@tempa \fi \ifx
  \@tempb \@empty \def\@tempb {arXiv}\fi \@ifundefined
  {mn@eprint@\@tempb}{\@tempb:\@tempc}{\expandafter \expandafter \csname
  mn@eprint@\@tempb\endcsname \expandafter{\@tempc}}}

\bibitem[\protect\citeauthoryear{Abdalla et~al.}{Abdalla
  et~al.}{2022}]{Abdalla:2022yfr}
Abdalla E.,  et~al., 2022, \mn@doi [JHEAp] {10.1016/j.jheap.2022.04.002}, 34,
  49

\bibitem[\protect\citeauthoryear{Abitbol et~al.}{Abitbol
  et~al.}{2019}]{SimonsObservatory:2019qwx}
Abitbol M.~H.,  et~al., 2019, Bull. Am. Astron. Soc., 51, 147

\bibitem[\protect\citeauthoryear{Addison, Watts, Bennett, Halpern, Hinshaw  \&
  Weiland}{Addison et~al.}{2018}]{Addison:2017fdm}
Addison G.~E.,  Watts D.~J.,  Bennett C.~L.,  Halpern M.,  Hinshaw G.,
  Weiland J.~L.,  2018, \mn@doi [Astrophys. J.] {10.3847/1538-4357/aaa1ed},
  853, 119

\bibitem[\protect\citeauthoryear{Ade et~al.}{Ade et~al.}{2014}]{Planck:2013nga}
Ade P. A.~R.,  et~al., 2014, \mn@doi [Astron. Astrophys.]
  {10.1051/0004-6361/201323003}, 566, A54

\bibitem[\protect\citeauthoryear{Ade et~al.}{Ade
  et~al.}{2019}]{SimonsObservatory:2018koc}
Ade P.,  et~al., 2019, \mn@doi [JCAP] {10.1088/1475-7516/2019/02/056}, 02, 056

\bibitem[\protect\citeauthoryear{Aghanim et~al.}{Aghanim
  et~al.}{2020a}]{Planck:2019nip}
Aghanim N.,  et~al., 2020a, \mn@doi [Astron. Astrophys.]
  {10.1051/0004-6361/201936386}, 641, A5

\bibitem[\protect\citeauthoryear{Aghanim et~al.}{Aghanim
  et~al.}{2020b}]{Planck:2018vyg}
Aghanim N.,  et~al., 2020b, \mn@doi [Astron. Astrophys.]
  {10.1051/0004-6361/201833910}, 641, A6

\bibitem[\protect\citeauthoryear{{Agrawal}, {Cyr-Racine}, {Pinner}  \&
  {Randall}}{{Agrawal} et~al.}{2019}]{Agrawal:2019lmo}
{Agrawal} P.,  {Cyr-Racine} F.-Y.,  {Pinner} D.,   {Randall} L.,  \href
  {https://ui.adsabs.harvard.edu/abs/2019arXiv190401016A} {arXiv:1904.01016}

\bibitem[\protect\citeauthoryear{Akaike}{Akaike}{1974}]{1100705}
Akaike H.,  1974, \mn@doi [IEEE Transactions on Automatic Control]
  {10.1109/TAC.1974.1100705}, 19, 716

\bibitem[\protect\citeauthoryear{Alam et~al.}{Alam et~al.}{2017}]{BOSS:2016wmc}
Alam S.,  et~al., 2017, \mn@doi [Mon. Not. Roy. Astron. Soc.]
  {10.1093/mnras/stx721}, 470, 2617

\bibitem[\protect\citeauthoryear{Alexander \& McDonough}{Alexander \&
  McDonough}{2019}]{Alexander:2019rsc}
Alexander S.,  McDonough E.,  2019, \mn@doi [Phys. Lett. B]
  {10.1016/j.physletb.2019.134830}, 797, 134830

\bibitem[\protect\citeauthoryear{Allali, Hertzberg  \& Rompineve}{Allali
  et~al.}{2021}]{Allali:2021azp}
Allali I.~J.,  Hertzberg M.~P.,   Rompineve F.,  2021, \mn@doi [Phys. Rev. D]
  {10.1103/PhysRevD.104.L081303}, 104, L081303

\bibitem[\protect\citeauthoryear{Archidiacono, Brinckmann, Lesgourgues  \&
  Poulin}{Archidiacono et~al.}{2017}]{Archidiacono:2016lnv}
Archidiacono M.,  Brinckmann T.,  Lesgourgues J.,   Poulin V.,  2017, \mn@doi
  [JCAP] {10.1088/1475-7516/2017/02/052}, 02, 052

\bibitem[\protect\citeauthoryear{Archidiacono, Hannestad  \&
  Lesgourgues}{Archidiacono et~al.}{2020}]{Archidiacono:2020dvx}
Archidiacono M.,  Hannestad S.,   Lesgourgues J.,  2020, \mn@doi [JCAP]
  {10.1088/1475-7516/2020/09/021}, 09, 021

\bibitem[\protect\citeauthoryear{Audren, Lesgourgues, Benabed  \&
  Prunet}{Audren et~al.}{2013}]{Audren:2012wb}
Audren B.,  Lesgourgues J.,  Benabed K.,   Prunet S.,  2013, \mn@doi [JCAP]
  {10.1088/1475-7516/2013/02/001}, 02, 001

\bibitem[\protect\citeauthoryear{Aylor, Joy, Knox, Millea, Raghunathan  \&
  Wu}{Aylor et~al.}{2019}]{Aylor:2018drw}
Aylor K.,  Joy M.,  Knox L.,  Millea M.,  Raghunathan S.,   Wu W. L.~K.,  2019,
  \mn@doi [Astrophys. J.] {10.3847/1538-4357/ab0898}, 874, 4

\bibitem[\protect\citeauthoryear{Ballesteros, Notari  \& Rompineve}{Ballesteros
  et~al.}{2020}]{Ballesteros:2020sik}
Ballesteros G.,  Notari A.,   Rompineve F.,  2020, \mn@doi [JCAP]
  {10.1088/1475-7516/2020/11/024}, 11, 024

\bibitem[\protect\citeauthoryear{Baumann, Nicolis, Senatore  \&
  Zaldarriaga}{Baumann et~al.}{2012}]{Baumann:2010tm}
Baumann D.,  Nicolis A.,  Senatore L.,   Zaldarriaga M.,  2012, \mn@doi [JCAP]
  {10.1088/1475-7516/2012/07/051}, 07, 051

\bibitem[\protect\citeauthoryear{Benevento, Kable, Addison  \&
  Bennett}{Benevento et~al.}{2022}]{Benevento:2022cql}
Benevento G.,  Kable J.~A.,  Addison G.~E.,   Bennett C.~L.,  2022, \mn@doi
  [Astrophys. J.] {10.3847/1538-4357/ac80fd}, 935, 156

\bibitem[\protect\citeauthoryear{Bernal, Verde  \& Riess}{Bernal
  et~al.}{2016}]{Bernal:2016gxb}
Bernal J.~L.,  Verde L.,   Riess A.~G.,  2016, \mn@doi [JCAP]
  {10.1088/1475-7516/2016/10/019}, 10, 019

\bibitem[\protect\citeauthoryear{Beutler \& McDonald}{Beutler \&
  McDonald}{2021}]{Beutler:2021eqq}
Beutler F.,  McDonald P.,  2021, \mn@doi [JCAP]
  {10.1088/1475-7516/2021/11/031}, 11, 031

\bibitem[\protect\citeauthoryear{Blas, Lesgourgues  \& Tram}{Blas
  et~al.}{2011}]{Blas:2011rf}
Blas D.,  Lesgourgues J.,   Tram T.,  2011, \mn@doi [JCAP]
  {10.1088/1475-7516/2011/07/034}, 07, 034

\bibitem[\protect\citeauthoryear{Boyle \& Komatsu}{Boyle \&
  Komatsu}{2018}]{Boyle:2017lzt}
Boyle A.,  Komatsu E.,  2018, \mn@doi [JCAP] {10.1088/1475-7516/2018/03/035},
  03, 035

\bibitem[\protect\citeauthoryear{Braglia, Ballardini, Emond, Finelli,
  Gumrukcuoglu, Koyama  \& Paoletti}{Braglia et~al.}{2020a}]{Braglia:2020iik}
Braglia M.,  Ballardini M.,  Emond W.~T.,  Finelli F.,  Gumrukcuoglu A.~E.,
  Koyama K.,   Paoletti D.,  2020a, \mn@doi [Phys. Rev. D]
  {10.1103/PhysRevD.102.023529}, 102, 023529

\bibitem[\protect\citeauthoryear{Braglia, Emond, Finelli, Gumrukcuoglu  \&
  Koyama}{Braglia et~al.}{2020b}]{Braglia:2020bym}
Braglia M.,  Emond W.~T.,  Finelli F.,  Gumrukcuoglu A.~E.,   Koyama K.,
  2020b, \mn@doi [Phys. Rev. D] {10.1103/PhysRevD.102.083513}, 102, 083513

\bibitem[\protect\citeauthoryear{Braglia, Ballardini, Finelli  \&
  Koyama}{Braglia et~al.}{2021}]{Braglia:2020auw}
Braglia M.,  Ballardini M.,  Finelli F.,   Koyama K.,  2021, \mn@doi [Phys.
  Rev. D] {10.1103/PhysRevD.103.043528}, 103, 043528

\bibitem[\protect\citeauthoryear{Brinckmann \& Lesgourgues}{Brinckmann \&
  Lesgourgues}{2019}]{Brinckmann:2018cvx}
Brinckmann T.,  Lesgourgues J.,  2019, \mn@doi [Phys. Dark Univ.]
  {10.1016/j.dark.2018.100260}, 24, 100260

\bibitem[\protect\citeauthoryear{Carrillo~Gonz\'alez, Liang, Sakstein  \&
  Trodden}{Carrillo~Gonz\'alez et~al.}{2021}]{CarrilloGonzalez:2020oac}
Carrillo~Gonz\'alez M.,  Liang Q.,  Sakstein J.,   Trodden M.,  2021, \mn@doi
  [JCAP] {10.1088/1475-7516/2021/04/063}, 04, 063

\bibitem[\protect\citeauthoryear{Chudaykin, Gorbunov  \& Nedelko}{Chudaykin
  et~al.}{2020a}]{Chudaykin:2020acu}
Chudaykin A.,  Gorbunov D.,   Nedelko N.,  2020a, \mn@doi [JCAP]
  {10.1088/1475-7516/2020/08/013}, 08, 013

\bibitem[\protect\citeauthoryear{Chudaykin, Ivanov, Philcox  \&
  Simonovi\'c}{Chudaykin et~al.}{2020b}]{Chudaykin:2020aoj}
Chudaykin A.,  Ivanov M.~M.,  Philcox O. H.~E.,   Simonovi\'c M.,  2020b,
  \mn@doi [Phys. Rev. D] {10.1103/PhysRevD.102.063533}, 102, 063533

\bibitem[\protect\citeauthoryear{{Chudaykin}, {Gorbunov}  \&
  {Nedelko}}{{Chudaykin} et~al.}{2022}]{Chudaykin:2022rnl}
{Chudaykin} A.,  {Gorbunov} D.,   {Nedelko} N.,  \href
  {https://ui.adsabs.harvard.edu/abs/2022arXiv220303666C} {arXiv:2203.03666}

\bibitem[\protect\citeauthoryear{{Clark}, {Vattis}, {Fan}  \&
  {Koushiappas}}{{Clark} et~al.}{2021}]{Clark:2021hlo}
{Clark} S.~J.,  {Vattis} K.,  {Fan} J.,   {Koushiappas} S.~M.,  \href
  {https://ui.adsabs.harvard.edu/abs/2021arXiv211009562C} {arXiv:2110.09562}

\bibitem[\protect\citeauthoryear{D'Amico, Senatore, Zhang  \& Zheng}{D'Amico
  et~al.}{2021}]{DAmico:2020ods}
D'Amico G.,  Senatore L.,  Zhang P.,   Zheng H.,  2021, \mn@doi [JCAP]
  {10.1088/1475-7516/2021/05/072}, 05, 072

\bibitem[\protect\citeauthoryear{Das, Maharana, Poulin  \& Sharma}{Das
  et~al.}{2022}]{Das:2021pof}
Das S.,  Maharana A.,  Poulin V.,   Sharma R.~K.,  2022, \mn@doi [Phys. Rev. D]
  {10.1103/PhysRevD.105.103503}, 105, 103503

\bibitem[\protect\citeauthoryear{Di~Valentino \& Bridle}{Di~Valentino \&
  Bridle}{2018}]{DiValentino:2018gcu}
Di~Valentino E.,  Bridle S.,  2018, \mn@doi [Symmetry] {10.3390/sym10110585},
  10, 585

\bibitem[\protect\citeauthoryear{Di~Valentino \& Melchiorri}{Di~Valentino \&
  Melchiorri}{2022}]{DiValentino:2021imh}
Di~Valentino E.,  Melchiorri A.,  2022, \mn@doi [Astrophys. J. Lett.]
  {10.3847/2041-8213/ac6ef5}, 931, L18

\bibitem[\protect\citeauthoryear{Di~Valentino et~al.,}{Di~Valentino
  et~al.}{2021}]{DiValentino:2021izs}
Di~Valentino E.,  et~al., 2021, \mn@doi [Class. Quant. Grav.]
  {10.1088/1361-6382/ac086d}, 38, 153001

\bibitem[\protect\citeauthoryear{{Efstathiou}}{{Efstathiou}}{2020}]{Efstathiou:2020wxn}
{Efstathiou} G.,  \href
  {https://ui.adsabs.harvard.edu/abs/2020arXiv200710716E} {arXiv:2007.10716}

\bibitem[\protect\citeauthoryear{Efstathiou}{Efstathiou}{2021}]{Efstathiou:2021ocp}
Efstathiou G.,  2021, \mn@doi [Mon. Not. Roy. Astron. Soc.]
  {10.1093/mnras/stab1588}, 505, 3866

\bibitem[\protect\citeauthoryear{Freedman et~al.}{Freedman
  et~al.}{2019}]{Freedman:2019jwv}
Freedman W.~L.,  et~al., 2019, \mn@doi [Astrophys. J.]
  {10.3847/1538-4357/ab2f73}, 882, 34

\bibitem[\protect\citeauthoryear{Freese \& Winkler}{Freese \&
  Winkler}{2021}]{Freese:2021rjq}
Freese K.,  Winkler M.~W.,  2021, \mn@doi [Phys. Rev. D]
  {10.1103/PhysRevD.104.083533}, 104, 083533

\bibitem[\protect\citeauthoryear{Gelman \& Rubin}{Gelman \&
  Rubin}{1992}]{Gelman-Rubin}
Gelman A.,  Rubin D.~B.,  1992, Statistical Science, 7, 457

\bibitem[\protect\citeauthoryear{Giusarma, Vagnozzi, Ho, Ferraro, Freese,
  Kamen-Rubio  \& Luk}{Giusarma et~al.}{2018}]{Giusarma:2018jei}
Giusarma E.,  Vagnozzi S.,  Ho S.,  Ferraro S.,  Freese K.,  Kamen-Rubio R.,
  Luk K.-B.,  2018, \mn@doi [Phys. Rev. D] {10.1103/PhysRevD.98.123526}, 98,
  123526

\bibitem[\protect\citeauthoryear{Gogoi, Sharma, Chanda  \& Das}{Gogoi
  et~al.}{2021}]{Gogoi:2020qif}
Gogoi A.,  Sharma R.~K.,  Chanda P.,   Das S.,  2021, \mn@doi [Astrophys. J.]
  {10.3847/1538-4357/abfe5b}, 915, 132

\bibitem[\protect\citeauthoryear{G\'omez-Valent}{G\'omez-Valent}{2022}]{Gomez-Valent:2022hkb}
G\'omez-Valent A.,  2022, \mn@doi [Phys. Rev. D] {10.1103/PhysRevD.106.063506},
  106, 063506

\bibitem[\protect\citeauthoryear{Guo, Zhang  \& Zhang}{Guo
  et~al.}{2019}]{Guo:2018ans}
Guo R.-Y.,  Zhang J.-F.,   Zhang X.,  2019, \mn@doi [JCAP]
  {10.1088/1475-7516/2019/02/054}, 02, 054

\bibitem[\protect\citeauthoryear{Handley \& Lemos}{Handley \&
  Lemos}{2021}]{Handley:2020hdp}
Handley W.,  Lemos P.,  2021, \mn@doi [Phys. Rev. D]
  {10.1103/PhysRevD.103.063529}, 103, 063529

\bibitem[\protect\citeauthoryear{{Herold} \& {Ferreira}}{{Herold} \&
  {Ferreira}}{2022}]{Herold:2022iib}
{Herold} L.,  {Ferreira} E. G.~M.,  \href
  {https://ui.adsabs.harvard.edu/abs/2022arXiv221016296H} {arXiv:2210.16296}

\bibitem[\protect\citeauthoryear{Herold, Ferreira  \& Komatsu}{Herold
  et~al.}{2022}]{Herold:2021ksg}
Herold L.,  Ferreira E. G.~M.,   Komatsu E.,  2022, \mn@doi [Astrophys. J.
  Lett.] {10.3847/2041-8213/ac63a3}, 929, L16

\bibitem[\protect\citeauthoryear{Hill, McDonough, Toomey  \& Alexander}{Hill
  et~al.}{2020}]{Hill:2020osr}
Hill J.~C.,  McDonough E.,  Toomey M.~W.,   Alexander S.,  2020, \mn@doi [Phys.
  Rev. D] {10.1103/PhysRevD.102.043507}, 102, 043507

\bibitem[\protect\citeauthoryear{Hill et~al.}{Hill et~al.}{2022}]{Hill:2021yec}
Hill J.~C.,  et~al., 2022, \mn@doi [Phys. Rev. D]
  {10.1103/PhysRevD.105.123536}, 105, 123536

\bibitem[\protect\citeauthoryear{Hou et~al.}{Hou et~al.}{2014}]{Hou:2012xq}
Hou Z.,  et~al., 2014, \mn@doi [Astrophys. J.] {10.1088/0004-637X/782/2/74},
  782, 74

\bibitem[\protect\citeauthoryear{Ili\'c, Sakr  \& Blanchard}{Ili\'c
  et~al.}{2019}]{Ilic:2019pwq}
Ili\'c S.,  Sakr Z.,   Blanchard A.,  2019, \mn@doi [Astron. Astrophys.]
  {10.1051/0004-6361/201936423}, 631, A96

\bibitem[\protect\citeauthoryear{Ivanov, Simonovi\'c  \& Zaldarriaga}{Ivanov
  et~al.}{2020a}]{Ivanov:2019pdj}
Ivanov M.~M.,  Simonovi\'c M.,   Zaldarriaga M.,  2020a, \mn@doi [JCAP]
  {10.1088/1475-7516/2020/05/042}, 05, 042

\bibitem[\protect\citeauthoryear{Ivanov, Simonovi\'c  \& Zaldarriaga}{Ivanov
  et~al.}{2020b}]{Ivanov:2019hqk}
Ivanov M.~M.,  Simonovi\'c M.,   Zaldarriaga M.,  2020b, \mn@doi [Phys. Rev. D]
  {10.1103/PhysRevD.101.083504}, 101, 083504

\bibitem[\protect\citeauthoryear{Ivanov, McDonough, Hill, Simonovi\'c, Toomey,
  Alexander  \& Zaldarriaga}{Ivanov et~al.}{2020c}]{Ivanov:2020ril}
Ivanov M.~M.,  McDonough E.,  Hill J.~C.,  Simonovi\'c M.,  Toomey M.~W.,
  Alexander S.,   Zaldarriaga M.,  2020c, \mn@doi [Phys. Rev. D]
  {10.1103/PhysRevD.102.103502}, 102, 103502

\bibitem[\protect\citeauthoryear{James \& Roos}{James \&
  Roos}{1975}]{James:1975dr}
James F.,  Roos M.,  1975, \mn@doi [Comput. Phys. Commun.]
  {10.1016/0010-4655(75)90039-9}, 10, 343

\bibitem[\protect\citeauthoryear{Jiang \& Piao}{Jiang \&
  Piao}{2021}]{Jiang:2021bab}
Jiang J.-Q.,  Piao Y.-S.,  2021, \mn@doi [Phys. Rev. D]
  {10.1103/PhysRevD.104.103524}, 104, 103524

\bibitem[\protect\citeauthoryear{Jiang \& Piao}{Jiang \&
  Piao}{2022}]{Jiang:2022uyg}
Jiang J.-Q.,  Piao Y.-S.,  2022, \mn@doi [Phys. Rev. D]
  {10.1103/PhysRevD.105.103514}, 105, 103514

\bibitem[\protect\citeauthoryear{{Jiang}, {Ye}  \& {Piao}}{{Jiang}
  et~al.}{2022}]{Jiang:2022qlj}
{Jiang} J.-Q.,  {Ye} G.,   {Piao} Y.-S., \href
  {https://ui.adsabs.harvard.edu/abs/2022arXiv221006125J} {arXiv:2210.06125}

\bibitem[\protect\citeauthoryear{Karwal \& Kamionkowski}{Karwal \&
  Kamionkowski}{2016}]{Karwal:2016vyq}
Karwal T.,  Kamionkowski M.,  2016, \mn@doi [Phys. Rev. D]
  {10.1103/PhysRevD.94.103523}, 94, 103523

\bibitem[\protect\citeauthoryear{Karwal, Raveri, Jain, Khoury  \&
  Trodden}{Karwal et~al.}{2022}]{Karwal:2021vpk}
Karwal T.,  Raveri M.,  Jain B.,  Khoury J.,   Trodden M.,  2022, \mn@doi
  [Phys. Rev. D] {10.1103/PhysRevD.105.063535}, 105, 063535

\bibitem[\protect\citeauthoryear{Khosravi \& Farhang}{Khosravi \&
  Farhang}{2022}]{Khosravi:2021csn}
Khosravi N.,  Farhang M.,  2022, \mn@doi [Phys. Rev. D]
  {10.1103/PhysRevD.105.063505}, 105, 063505

\bibitem[\protect\citeauthoryear{Klypin et~al.,}{Klypin
  et~al.}{2021}]{Klypin:2020tud}
Klypin A.,  et~al., 2021, \mn@doi [Mon. Not. Roy. Astron. Soc.]
  {10.1093/mnras/stab769}, 504, 769

\bibitem[\protect\citeauthoryear{Knox \& Millea}{Knox \&
  Millea}{2020}]{Knox:2019rjx}
Knox L.,  Millea M.,  2020, \mn@doi [Phys. Rev. D]
  {10.1103/PhysRevD.101.043533}, 101, 043533

\bibitem[\protect\citeauthoryear{Krishnan, Colg\'ain, Ruchika, Sen,
  Sheikh-Jabbari  \& Yang}{Krishnan et~al.}{2020}]{Krishnan:2020obg}
Krishnan C.,  Colg\'ain E.~O.,  Ruchika Sen A.~A.,  Sheikh-Jabbari M.~M.,
  Yang T.,  2020, \mn@doi [Phys. Rev. D] {10.1103/PhysRevD.102.103525}, 102,
  103525

\bibitem[\protect\citeauthoryear{La~Posta, Louis, Garrido  \& Hill}{La~Posta
  et~al.}{2022}]{LaPosta:2021pgm}
La~Posta A.,  Louis T.,  Garrido X.,   Hill J.~C.,  2022, \mn@doi [Phys. Rev.
  D] {10.1103/PhysRevD.105.083519}, 105, 083519

\bibitem[\protect\citeauthoryear{Lattanzi \& Gerbino}{Lattanzi \&
  Gerbino}{2018}]{Lattanzi:2017ubx}
Lattanzi M.,  Gerbino M.,  2018, \mn@doi [Front. in Phys.]
  {10.3389/fphy.2017.00070}, 5, 70

\bibitem[\protect\citeauthoryear{Lemos, Lee, Efstathiou  \& Gratton}{Lemos
  et~al.}{2019}]{Lemos:2018smw}
Lemos P.,  Lee E.,  Efstathiou G.,   Gratton S.,  2019, \mn@doi [Mon. Not. Roy.
  Astron. Soc.] {10.1093/mnras/sty3082}, 483, 4803

\bibitem[\protect\citeauthoryear{Lesgourgues \& Pastor}{Lesgourgues \&
  Pastor}{2006}]{Lesgourgues:2006nd}
Lesgourgues J.,  Pastor S.,  2006, \mn@doi [Phys. Rept.]
  {10.1016/j.physrep.2006.04.001}, 429, 307

\bibitem[\protect\citeauthoryear{Lin, Benevento, Hu  \& Raveri}{Lin
  et~al.}{2019}]{Lin:2019qug}
Lin M.-X.,  Benevento G.,  Hu W.,   Raveri M.,  2019, \mn@doi [Phys. Rev. D]
  {10.1103/PhysRevD.100.063542}, 100, 063542

\bibitem[\protect\citeauthoryear{M\"ortsell \& Dhawan}{M\"ortsell \&
  Dhawan}{2018}]{Mortsell:2018mfj}
M\"ortsell E.,  Dhawan S.,  2018, \mn@doi [JCAP]
  {10.1088/1475-7516/2018/09/025}, 09, 025

\bibitem[\protect\citeauthoryear{Mortsell, Goobar, Johansson  \&
  Dhawan}{Mortsell et~al.}{2022}]{Mortsell:2021nzg}
Mortsell E.,  Goobar A.,  Johansson J.,   Dhawan S.,  2022, \mn@doi [Astrophys.
  J.] {10.3847/1538-4357/ac756e}, 933, 212

\bibitem[\protect\citeauthoryear{Murgia, Abell\'an  \& Poulin}{Murgia
  et~al.}{2021}]{Murgia:2020ryi}
Murgia R.,  Abell\'an G.~F.,   Poulin V.,  2021, \mn@doi [Phys. Rev. D]
  {10.1103/PhysRevD.103.063502}, 103, 063502

\bibitem[\protect\citeauthoryear{Niedermann \& Sloth}{Niedermann \&
  Sloth}{2021}]{Niedermann:2019olb}
Niedermann F.,  Sloth M.~S.,  2021, \mn@doi [Phys. Rev. D]
  {10.1103/PhysRevD.103.L041303}, 103, L041303

\bibitem[\protect\citeauthoryear{Niedermann \& Sloth}{Niedermann \&
  Sloth}{2022}]{Niedermann:2021vgd}
Niedermann F.,  Sloth M.~S.,  2022, \mn@doi [Phys. Rev. D]
  {10.1103/PhysRevD.105.063509}, 105, 063509

\bibitem[\protect\citeauthoryear{Nojiri, Odintsov, Saez-Chillon~Gomez  \&
  Sharov}{Nojiri et~al.}{2021}]{Nojiri:2021dze}
Nojiri S.,  Odintsov S.~D.,  Saez-Chillon~Gomez D.,   Sharov G.~S.,  2021,
  \mn@doi [Phys. Dark Univ.] {10.1016/j.dark.2021.100837}, 32, 100837

\bibitem[\protect\citeauthoryear{Nunes \& Vagnozzi}{Nunes \&
  Vagnozzi}{2021}]{Nunes:2021ipq}
Nunes R.~C.,  Vagnozzi S.,  2021, \mn@doi [Mon. Not. Roy. Astron. Soc.]
  {10.1093/mnras/stab1613}, 505, 5427

\bibitem[\protect\citeauthoryear{Nunes, Vagnozzi, Kumar, Di~Valentino  \&
  Mena}{Nunes et~al.}{2022}]{Nunes:2022bhn}
Nunes R.~C.,  Vagnozzi S.,  Kumar S.,  Di~Valentino E.,   Mena O.,  2022,
  \mn@doi [Phys. Rev. D] {10.1103/PhysRevD.105.123506}, 105, 123506

\bibitem[\protect\citeauthoryear{Oikonomou}{Oikonomou}{2021}]{Oikonomou:2020qah}
Oikonomou V.~K.,  2021, \mn@doi [Phys. Rev. D] {10.1103/PhysRevD.103.044036},
  103, 044036

\bibitem[\protect\citeauthoryear{Perivolaropoulos \& Skara}{Perivolaropoulos \&
  Skara}{2022}]{Perivolaropoulos:2021jda}
Perivolaropoulos L.,  Skara F.,  2022, \mn@doi [New Astron. Rev.]
  {10.1016/j.newar.2022.101659}, 95, 101659

\bibitem[\protect\citeauthoryear{Philcox, Ivanov, Simonovi\'c  \&
  Zaldarriaga}{Philcox et~al.}{2020}]{Philcox:2020vvt}
Philcox O. H.~E.,  Ivanov M.~M.,  Simonovi\'c M.,   Zaldarriaga M.,  2020,
  \mn@doi [JCAP] {10.1088/1475-7516/2020/05/032}, 05, 032

\bibitem[\protect\citeauthoryear{Poulin, Smith, Karwal  \& Kamionkowski}{Poulin
  et~al.}{2019}]{Poulin:2018cxd}
Poulin V.,  Smith T.~L.,  Karwal T.,   Kamionkowski M.,  2019, \mn@doi [Phys.
  Rev. Lett.] {10.1103/PhysRevLett.122.221301}, 122, 221301

\bibitem[\protect\citeauthoryear{Poulin, Smith  \& Bartlett}{Poulin
  et~al.}{2021}]{Poulin:2021bjr}
Poulin V.,  Smith T.~L.,   Bartlett A.,  2021, \mn@doi [Phys. Rev. D]
  {10.1103/PhysRevD.104.123550}, 104, 123550

\bibitem[\protect\citeauthoryear{Riess et~al.}{Riess
  et~al.}{2022}]{Riess:2021jrx}
Riess A.~G.,  et~al., 2022, \mn@doi [Astrophys. J. Lett.]
  {10.3847/2041-8213/ac5c5b}, 934, L7

\bibitem[\protect\citeauthoryear{Roy~Choudhury \& Hannestad}{Roy~Choudhury \&
  Hannestad}{2020}]{RoyChoudhury:2019hls}
Roy~Choudhury S.,  Hannestad S.,  2020, \mn@doi [JCAP]
  {10.1088/1475-7516/2020/07/037}, 07, 037

\bibitem[\protect\citeauthoryear{Sabla \& Caldwell}{Sabla \&
  Caldwell}{2022}]{Sabla:2022xzj}
Sabla V.~I.,  Caldwell R.~R.,  2022, \mn@doi [Phys. Rev. D]
  {10.1103/PhysRevD.106.063526}, 106, 063526

\bibitem[\protect\citeauthoryear{Sakr}{Sakr}{2022}]{Sakr:2022ans}
Sakr Z.,  2022, \mn@doi [Universe] {10.3390/universe8050284}, 8, 284

\bibitem[\protect\citeauthoryear{Sakr, Ilic  \& Blanchard}{Sakr
  et~al.}{2022}]{Sakr:2021jya}
Sakr Z.,  Ilic S.,   Blanchard A.,  2022, \mn@doi [Astron. Astrophys.]
  {10.1051/0004-6361/202142115}, 666, A34

\bibitem[\protect\citeauthoryear{Sakstein \& Trodden}{Sakstein \&
  Trodden}{2020}]{Sakstein:2019fmf}
Sakstein J.,  Trodden M.,  2020, \mn@doi [Phys. Rev. Lett.]
  {10.1103/PhysRevLett.124.161301}, 124, 161301

\bibitem[\protect\citeauthoryear{Sch\"oneberg, Franco~Abell\'an,
  P\'erez~S\'anchez, Witte, Poulin  \& Lesgourgues}{Sch\"oneberg
  et~al.}{2022}]{Schoneberg:2021qvd}
Sch\"oneberg N.,  Franco~Abell\'an G.,  P\'erez~S\'anchez A.,  Witte S.~J.,
  Poulin V.,   Lesgourgues J.,  2022, \mn@doi [Phys. Rept.]
  {10.1016/j.physrep.2022.07.001}, 984, 1

\bibitem[\protect\citeauthoryear{Smith, Poulin  \& Amin}{Smith
  et~al.}{2020}]{Smith:2019ihp}
Smith T.~L.,  Poulin V.,   Amin M.~A.,  2020, \mn@doi [Phys. Rev. D]
  {10.1103/PhysRevD.101.063523}, 101, 063523

\bibitem[\protect\citeauthoryear{Smith, Poulin, Bernal, Boddy, Kamionkowski  \&
  Murgia}{Smith et~al.}{2021}]{Smith:2020rxx}
Smith T.~L.,  Poulin V.,  Bernal J.~L.,  Boddy K.~K.,  Kamionkowski M.,
  Murgia R.,  2021, \mn@doi [Phys. Rev. D] {10.1103/PhysRevD.103.123542}, 103,
  123542

\bibitem[\protect\citeauthoryear{Smith, Lucca, Poulin, Abellan, Balkenhol,
  Benabed, Galli  \& Murgia}{Smith et~al.}{2022}]{Smith:2022hwi}
Smith T.~L.,  Lucca M.,  Poulin V.,  Abellan G.~F.,  Balkenhol L.,  Benabed K.,
   Galli S.,   Murgia R.,  2022, \mn@doi [Phys. Rev. D]
  {10.1103/PhysRevD.106.043526}, 106, 043526

\bibitem[\protect\citeauthoryear{Tanseri, Hagstotz, Vagnozzi, Giusarma  \&
  Freese}{Tanseri et~al.}{2022}]{Tanseri:2022zfe}
Tanseri I.,  Hagstotz S.,  Vagnozzi S.,  Giusarma E.,   Freese K.,  2022,
  \mn@doi [JHEAp] {10.1016/j.jheap.2022.07.002}, 36, 1

\bibitem[\protect\citeauthoryear{{Vagnozzi}}{{Vagnozzi}}{2019}]{Vagnozzi:2019utt}
{Vagnozzi} S.,  \href
  {https://ui.adsabs.harvard.edu/abs/2019arXiv190708010V} {arXiv:1907.08010}

\bibitem[\protect\citeauthoryear{Vagnozzi}{Vagnozzi}{2020}]{Vagnozzi:2019ezj}
Vagnozzi S.,  2020, \mn@doi [Phys. Rev. D] {10.1103/PhysRevD.102.023518}, 102,
  023518

\bibitem[\protect\citeauthoryear{Vagnozzi}{Vagnozzi}{2021}]{Vagnozzi:2021gjh}
Vagnozzi S.,  2021, \mn@doi [Phys. Rev. D] {10.1103/PhysRevD.104.063524}, 104,
  063524

\bibitem[\protect\citeauthoryear{Vagnozzi, Giusarma, Mena, Freese, Gerbino, Ho
  \& Lattanzi}{Vagnozzi et~al.}{2017}]{Vagnozzi:2017ovm}
Vagnozzi S.,  Giusarma E.,  Mena O.,  Freese K.,  Gerbino M.,  Ho S.,
  Lattanzi M.,  2017, \mn@doi [Phys. Rev. D] {10.1103/PhysRevD.96.123503}, 96,
  123503

\bibitem[\protect\citeauthoryear{Vagnozzi, Dhawan, Gerbino, Freese, Goobar  \&
  Mena}{Vagnozzi et~al.}{2018}]{Vagnozzi:2018jhn}
Vagnozzi S.,  Dhawan S.,  Gerbino M.,  Freese K.,  Goobar A.,   Mena O.,  2018,
  \mn@doi [Phys. Rev. D] {10.1103/PhysRevD.98.083501}, 98, 083501

\bibitem[\protect\citeauthoryear{Ye \& Piao}{Ye \& Piao}{2020}]{Ye:2020btb}
Ye G.,  Piao Y.-S.,  2020, \mn@doi [Phys. Rev. D]
  {10.1103/PhysRevD.101.083507}, 101, 083507

\bibitem[\protect\citeauthoryear{{Ye}, {Zhang}  \& {Piao}}{{Ye}
  et~al.}{2021}]{Ye:2021iwa}
{Ye} G.,  {Zhang} J.,   {Piao} Y.-S., \href
  {https://ui.adsabs.harvard.edu/abs/2021arXiv210713391Y} {arXiv:2107.13391}

\bibitem[\protect\citeauthoryear{Ye, Jiang  \& Piao}{Ye
  et~al.}{2022}]{Ye:2022efx}
Ye G.,  Jiang J.-Q.,   Piao Y.-S.,  2022, \mn@doi [Phys. Rev. D]
  {10.1103/PhysRevD.106.103528}, 106, 103528

\bibitem[\protect\citeauthoryear{Zumalacarregui}{Zumalacarregui}{2020}]{Zumalacarregui:2020cjh}
Zumalacarregui M.,  2020, \mn@doi [Phys. Rev. D] {10.1103/PhysRevD.102.023523},
  102, 023523

\makeatother
\end{thebibliography}

\bsp

\label{lastpage}

\end{document}